\documentclass[draft]{agujournal2019}
\usepackage{amsmath,amsthm,amsfonts,mathtools,bm}
\newcommand{\mean}{\mu}
\newcommand{\m}{m}
\newcommand{\ta}{\m}
\newcommand{\y}{d}
\newcommand{\py}{\tilde{\y}}
\newcommand{\Y}{D}
\newcommand{\G}{G}
\newcommand{\w}{\omega}
\newcommand{\ww}{w}

\newcommand{\N}{\mathcal{N}}
\newcommand{\pMean}{\tilde{\mean}}
\newcommand{\pC}{\tilde{C}}

\usepackage{algorithm} 
\usepackage{algpseudocode}
\usepackage[english]{babel}

\usepackage[inline]{trackchanges} 
\usepackage{soul}
\usepackage{blindtext}
\usepackage{array}
\usepackage{multicol}
\usepackage{graphicx}

\graphicspath{ {./Figure/}}

\draftfalse
\journalname{XXXXXXXXX}

\begin{document}
\title{Multi-task unscented Kalman inversion (MUKI): a derivative-free joint inversion framework and its application to joint inversion of geophysical data}

%
%

\authors{Longlong Wang\affil{1,2}, Yun Chen\affil{1,3}, Youshan Liu\affil{1}, Nanqiao Du\affil{4},Wei Li\affil{5}, Junliu Suwen\affil{6}}
\affiliation{1}{State Key Laboratory of Lithospheric Evolution, Institute of Geology and Geophysics, Chinese Academy of Sciences}
\affiliation{2}{College of Earth and Planetary Sciences, University of Chinese Academy of Sciences, Beijing, China}
\affiliation{3}{CAS Center for Excellence in Deep Earth Science, Guangzhou, China}
\affiliation{4}{Department of Earth Science, University of Toronto, Toronto, Canada}
\affiliation{5}{Hubei Subsurface Multi‐Scale Imaging Key Laboratory, Institute of Geophysics and Geomatics, China University of Geosciences, Wuhan, China}
\affiliation{6}{Key Laboratory of Earth and Planetary Physics, Institute of Geology and Geophysics, Chinese Academy of Sciences, Beijing, China}

\correspondingauthor{Wang Longlong}{wanglonglong@mail.iggcas.ac.cn}
\correspondingauthor{Chen Yun}{yunchen@mail.iggcas.ac.cn}

\begin{keypoints}
\item Multi-task unscented Kalman inversion is developed based on the unscented Kalman inversion.
\item Multi-task unscented Kalman inversion is a derivative-free joint inversion framework and can be applied to the joint inversion of multi-physical data.
\item Tests of the joint inversion of receiver functions and surface wave demonstrate MUKI’s fast convergence rate has fast-speed convergences with uncertainty estimation analysics.
\end{keypoints}


\begin{abstract}
In the geophysical joint inversion, the gradient and Bayesian Markov Chain Monte Carlo (MCMC) sampling-based methods are widely used owing to their fast convergences or global optimality. However, these methods either require the computation of gradients and easily fall into local optimal solutions, or cost much time to carry out the millions of forward calculations in a huge sampling space. Different from these two methods, taking advantage of the recently developed unscented Kalman method in computational mathematics, we extend an iterative gradient-free Bayesian joint inversion framework, i.e., Multi-task unscented Kalman inversion (MUKI). In this new framework, information from various observations is incorporated, the model is iteratively updated in a derivative-free way, and a Gaussian approximation to the posterior distribution of the model parameters is obtained. We apply the MUKI to the joint inversion of receiver functions and surface wave dispersion, which is well-established and widely used to construct the crustal and upper mantle structure of the earth. Based on synthesized and real data, the tests demonstrate that MUKI can recover the model more efficiently than the gradient-based method and the Markov Chain Monte Carlo method, and it would be a promising approach to resolve the geophysical joint inversion problems.
\end{abstract}

\section*{Plain Language Summary}
There are lots of gradient-based optimization methods and Bayesian sampling methods used in joint inversion problems. However, each of them has some disadvantages. Gradient-based methods are usually efficient in fast convergence, but they easily fall into local minima; Bayesian sampling methods can help to seek global solutions but are computationally intensive. To take their advantages simultaneously, here we propose a new joint inversion framework, Multi-task unscented Kalman inversion (MUKI), based on the recently developed unscented Kalman inversion method. In this new framework, a gradient-free, iterative manner is adopted to yield a fast convergence, and a Bayesian approach is used for uncertainty analysis. Joint inversion of receiver functions and surface wave dispersion is well-established in geophysics and has been successfully applied with gradient-based methods or Bayesian sampling methods to reconstruct the structure of the earth’s crust and upper mantle. Thus, we use the joint inversion of receiver functions and surface wave as a benchmark for our newly developed MUKI. The tests of synthesized and real data prove that MUKI is more robust, accurate, and efficient compared to the gradient and Bayesian sampling-based methods.

\section{Introduction}
Geophysical inversion is an optimization problem that uses mathematical methods to estimate geological model parameters from a series of observed geophysical data \cite{Fichtner2021,Stuart2010,Tarantola2005}. Independent inversions of a single property are ineffective when the resolution of a single data set is limited. Inverting one geophysical data set with other supplemental geophysical data sets simultaneously, i.e., joint inversion, can improve inversion results by utilizing different sensitivities of different data sets to specific parameters and can mitigate the nonuniqueness of inversion problems compared with only using a single data set. 

Joint inversion can be divided into two groups, either multiple data sets that are sensitive to the same physical properties or different data sets that are responses to the different physical properties of the same geological structure~\cite{Moorkamp2007}. These two groups can both be solved by gradient-based methods, such as the damped least squares method~\cite{Gallardo2007,Julia2000}. Although gradient-based methods can be effective, these methods easily suffer from local minimum except for a good initial model. To obtain geological realistic models, these methods usually adopt some damping and smooth regularization, which may reduce accuracy or resolution. The recently developed Bayesian-based MCMC method can not only obtain a global optimal solution but also obtain uncertainties of results. However, the MCMC method is costly because millions of forward calculations are required to thoroughly sample the posterior probability density function. 

Overall, combining different model parameters with observed data sets for an effective joint inversion method remains a challenge. To solve this problem, we resort to a new inversion framework, i.e., Kalman filter-based (KF) inversion. KF was first proposed to describe a recursive solution to the discrete-data linear filtering problem in 1960~\cite{Kalman1960}. Since then, KF has been a subject of extensive research and application, particularly in the area of signal processing~\cite{Dean1986,Jwo2008,Plett2006}.

Two Kalman-based inversion methods can be applied to nonlinear systems, ensemble Kalman inversion (EnKI) and unscented Kalman inversion (UKI), which can be seen as applications of the corresponding filters in the optimization domain. KF itself is intended for linear systems. When the system is nonlinear, this method can be extended to the extended Kalman filter (ExKF) by linearizing the nonlinear system. However, the ExKF method is only reliable for weakly nonlinear problems. To resolve major problems of the ExKF for parameter estimation for strongly nonlinear models and large state spaces, \citeA{Evensen1994} proposed a Monte Carlo-based KF, i.e., ensemble Kalman filter. By using a collection of hundreds to thousands of state vectors to evaluate the system, this novel filter works well in both linear and nonlinear systems~\cite{Evensen2009,Iglesias2012}, and there are several optimization implementations in recent years~\cite{Aleardi2021,Conjard2021,Muir2019,Wang2019}.

Analogous to EnKI, UKI is also gradient-free but adopts some integration points (which are the deterministic sample points to capture the mean and covariance of the Gaussian distribution) to evaluate a nonlinear system. By using the so-called unscented transformation~\cite{Julier1995}, UKI has been shown to produce superior results to EnKI on a variety of nonlinear inverse problems~\cite{Huang2022}.

Here we extend UKI to multi-task UKI (MUKI) by introducing a generalized covariance function across different data sets and model parameters. As a benchmark, we apply the MUKI to the joint inversion of receiver functions and surface wave dispersion, which is a powerful tool to invert the crustal and upper mantle structure of the earth, and the gradient-based method and MCMC method have been successfully intro\cite{Julia2003,Sambridge2013}. The remainder of this paper will introduce and evaluate our developed MUKI method comprehensively against these classic methods.

\section{Method} \label{Method}
\subsection{Gaussian Approximation Algorithm} \label{GAA}
In general, geophysical inverse problems can be formulated as follows
\begin{equation}
\y = \G(\ta) + \eta,
\end{equation}  
where $\G:\mathbb{R}^{N_\ta}\rightarrow \mathbb{R}^{N_\y}$ is a nonlinear function (especially, a forward map in RF or SWD) that denotes a forward operator.
In inversion, the data $d$ is available, thus the inverse problem is to recover parameter vector $\ta \in \mathbb{R}^{N_\ta}$ from $\y \in \mathbb{R}^{N_y}$,
with the observed noise $\eta$ drawn from a Gaussian distribution $\mathcal{N}(0,\Sigma_{\eta})$.   

In the Bayesian viewpoint, the inverse process aims to maximize a posterior estimation for $\ta$, which can be written as

\begin{equation}
    p(\ta|\Y) \propto \exp (-\Phi(\ta)) p_{0}(\ta)\hspace{0.1in}\text { with } \hspace{0.1in} \Phi(\ta)=\frac{1}{2}\left\|\Sigma_{\eta}^{-\frac{1}{2}}(\y-\G(\ta))\right\|^{2}.
\end{equation}

Generally, the exact posterior distribution $p(\ta|\Y)$, is intractable to compute. In this paper, we view the process of maximizing a posterior estimation as a stochastic dynamic for the parameter, so that we can employ KF to estimate the parameter for a given observation~\cite{Huang2021a}.

Consider the stochastic dynamic system:

\begin{align}
&evolution:  \ta_{n+1} = \ta_{n} +  \omega_{n+1}, \omega_{n+1} \sim \N(0,\Sigma_{\omega}),\\
&observation:  \y_{n+1} = \G(\ta_{n+1}) + \nu_{n+1}, \nu_{n+1} \sim \N(0,\Sigma_{\nu}),
\end{align}
where the artificial evolution error covariance $\Sigma_\omega$ and the artificial observation error covariance $\Sigma_\nu$ are both positive definite matrices. We use the Gaussian Approximation Algorithm (GAA) to obtain $\m_n$ from $D_n={\y_1, \y_2,... \y_n}$, where $\y_n$ is the observation $d$ at $n$ iteration. Note that when we apply KF in optimization, the $\y_n$ is identical to $\y_{obs}$~\cite{Huang2022}.

The GAA, originally in geostatistics, also known as Gaussian process regression~\cite{Chiles2012}, is a method of interpolation that maps Gaussian distribution to Gaussian distribution
during a Gaussian process and leads to an insight into the Kalman methodology~\cite{Huang2022}. This algorithm starts with a Gaussian approximation to the posterior distribution
$p_0 \sim \mathcal{N}(\mean_0,C_0)$, $\mu_0$ and $C_0$ are the mean and covariance of the Gaussian distribution $p_0$, respectively. 
Then successive estimates, $p_1, p_2,... p_n$ can be iteratively  computed by repeatedly applying the following two steps, $p_n \mapsto \hat{p}_{n+1}$ and then $\hat{p}_{n+1} \mapsto p_{n+1}$. 
In the first step (analogue to prediction step in Kalman Filter), the predicted model parameters vector $\hat{p}_{n+1} \sim \mathcal{N}\left(\pMean_{n+1}, \pC_{n+1}\right)$ is also Gaussian and satisfies

\begin{eqnarray}
    \pMean_{n+1}&=&\mean_n,\label{pM}\\
    \pC_{n+1}&=&C_n+\Sigma_\omega,\label{pC}
\end{eqnarray}
where $\Sigma_\omega$ are observation error covariances, and $\tilde{.}$ denotes the updated parameters.
Then in the second step (correction step), we use the mean and covariance matrix to represent the joint distribution of $\left\{\ta_{n+1}, \y_{n+1} \mid \Y_{n}\right\}$,
which is a multivariate Gaussian distribution
\begin{eqnarray}\label{eq22}\mathcal{N}\left(\left[
\begin{array}{lcl}\pMean_{n+1}\\\py_{n+1}\end{array}\right]
,
\left[
\begin{array}{lcl}
\pC_{n+1} & \pC^{\ta d}_{n+1}\\ 
\pC^{d \ta}_{n+1} & \pC^{dd}_{n+1}
\end{array}
\right]
\right),\label{mGd}
\end{eqnarray}
where
\begin{eqnarray}
\py_{n+1}&=&\mathbb{E}[\y_{n+1}|\Y_n]=\mathbb{E}[\G(\ta_{n+1})|\Y_n], \label{YY}\\
\pC^{\ta d}_{n+1}&=&Cov[\ta_{n+1},\y_{n+1}|\Y_n]=Cov[\ta_{n+1},\G(\ta_{n+1})|\Y_n], \label{Cmp}\\
\pC^{dd}_{n+1}&=&Cov[\y_{n+1}|\Y_n]=Cov[\G(\ta_{n+1})|\Y_n] + \Sigma_\nu. \label{Cmm}
\end{eqnarray}
$\Sigma_\nu$ is the artificial observation error covariance.

Conditioning the joint Gaussian distribution in equation (\ref{eq22})
to obtain p($\ta_{n+1} | \Y_{n+1}=\y_{n+1})$~\cite{Bishop2006}, specific operations can also be found in Supplementary Text S1.
\begin{eqnarray}
\mean_{n+1}&=&\pMean_{n+1} + \pC^{\ta d}_{n+1}(\pC^{dd}_{n+1})^{-1}(\y_{n+1}-\py_{n+1}), \label{mn+1}  \\
C_{n+1}&=&\pC_{n+1} - \pC^{\ta d}_{n+1}(\pC^{dd}_{n+1})^{-1}\pC
^{\ta d}_{n+1}{}^T. \label{cn+1}
\end{eqnarray}

We can see that GAA not only provides the expected value $\mean_{n+1}$ for the given observation $\y_{n+1}$, 
but also offers uncertainty through the variance $C_{n+1}$. From an optimization point of view, 
the observation $D_n$ is not changed, it is identical to $d_{obs}$ at each iteration. 
By implementing approximations to equations (\ref{YY}) to (\ref{Cmm}), one iteration of the Kalman inversion will be established.

In this research, the parameters in equations (\ref{pC}), (\ref{Cmm}) are chosen as $\Sigma_{\w} = C_n$ and $\Sigma_{v}=2\Sigma_{\eta}$, 
which guarantees that the algorithm can obtain an accurate Gaussian estimation to the posterior probability with the
converged mean and covariance~\cite{Huang2021a}.

\subsection{Unscented Kalman Inversion (UKI)}
In the Kalman method, a vital operation is how to measure the
statistical properties of the system states, i.e. approximating
the equations (\ref{Cmp}),(\ref{Cmm}). The UKI method approximates the integrals by deterministic quadrature rules, which are generally 
called unscented transform~\cite{Julier1995}. The approach is also described in Supplementary Text S2. To evaluate the integration, 2$N_\ta+1$ ($N_m$ is the dimension of the vector $m$) integration points have been selected deterministically according to the posterior distribution $\mathcal{N}(\mean_{n+1}, C_{n+1})$

\begin{eqnarray}
\ta_{n+1}^j &=& \mean_{n+1}+c_j[\sqrt{C_{n+1}}]_j \hspace{0.2in} 1\le j \le N_\ta,\\
\ta_{n+1}^{j+N_\ta}  &=& \mean_{n+1}-c_j[\sqrt{C_{n+1}}]_j \hspace{0.2in} 1\le j \le N_\ta.
\end{eqnarray}
where $[\sqrt{C_{n+1}}]_j$ is the j-th column of the Cholesky factor of $C_{n+1}$ in the $n+1$ step, according to the parameter setting
in unscented transform, the coefficient $c_j=a\sqrt{N_\ta}$ and the hyperparameters $a= min\{\sqrt{\frac{4}{N_\ta}},1\}$. 
The central integration point $\ta_{n+1}^0=\mean_{n+1}$. 

The equations (\ref{YY}) to (\ref{Cmm}) can be defined as
\begin{eqnarray}
\py_{n+1}&=&\G(\ta_{n+1}^0), \\
\pC^{\ta\y}_{n+1}&=& \sum_{j=1}^{2N_\ta}\ww_j^c(\ta_{n+1}^j-\mean_{n+1})(\ta_{n+1}^j-\mathbb{E}\G(\mean_{n+1}))^T, \\
\pC^{\y\y}_{n+1}&=& \sum_{j=1}^{2N_\ta}\ww_j^c(\G(\ta_{n+1}^j)-\mathbb{E}\G(\mean_{n+1}))(\G(\ta_{n+1}^j)-\mathbb{E}\G(\mean_{n+1}))^T,
\end{eqnarray}
where quadrature weights $\ww_j^c = \frac{1}{2a^2N_\ta}$. Details on this approach can be found in \citeA{Huang2021a} and the supplementary Text S2.

\begin{figure}[htp]
    \noindent\includegraphics[width=\textwidth]{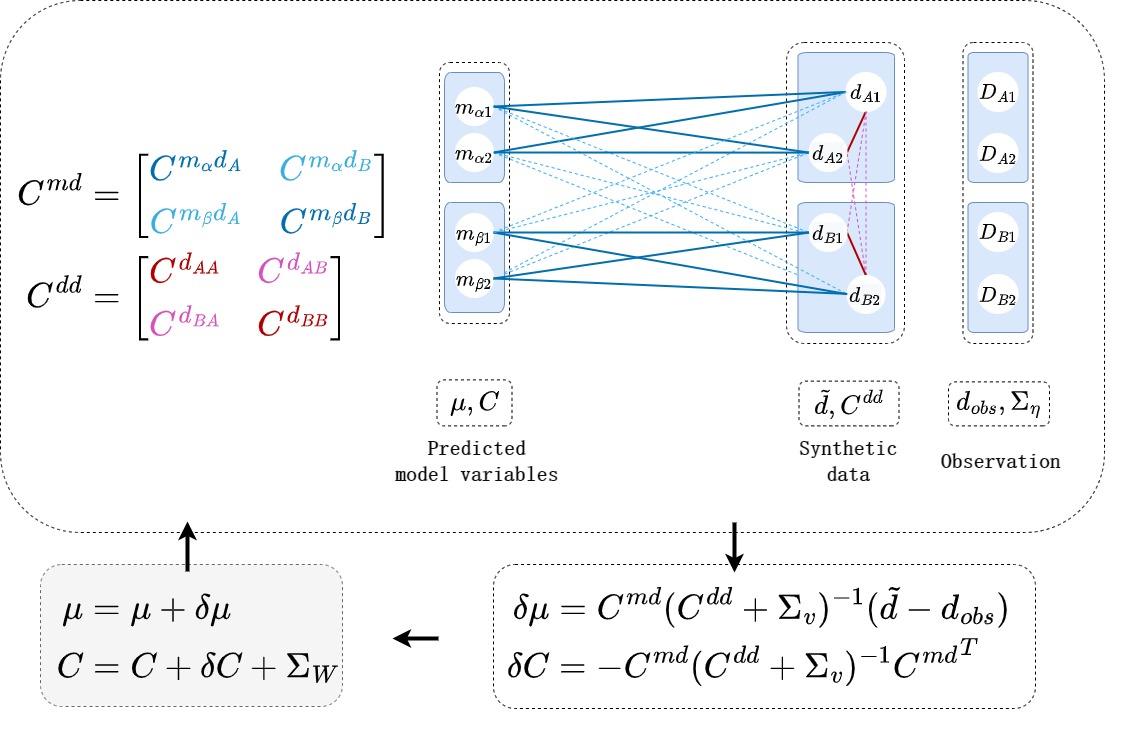}
    \caption{Schematical map of the MUKI framework. This diagram only shows the process of one model parameter update, the loop termination condition is not drawn and can be set according to the error function.}
    \label{a1}
\end{figure}

\subsection{ Multi-task unscented Kalman
inversion }
Inspired by the multi-task learning approach in machine learning problems~\cite{Bonilla2007}, in this study, we combine the observation data $\y_{A},\y_{B},\y_{C},...$~(i.e., the multi-responses of models), into a generalized $\y$, so that different observations (we only use $\y_A,\y_B,\y_C$ to demonstratively represent multi-responses of models in the remaining part of the paper) can share covariance.
The generalized $\y$ is set as
\begin{eqnarray}
\y = [\alpha\y_{A1},...,\alpha\y_{A_m},\beta\y_{B_1},...,\beta\y_{B_n},\gamma\y_{C1},...,\gamma\y_{Co},...]^T,
\end{eqnarray}
where $\alpha,\beta,\gamma,...$ are custom weight factors according to the relative quality of data and weight. 

We can summarize the process of the MUKI inversion in Figure 1. The process shown in Figure 1 can be divided into two steps, the prediction step (grey background) and the correction step (white background), which correspond to the equations (\ref{pM}) to (\ref{pC}) and (\ref{Cmp})-(\ref{cn+1}), respectively. In the correction step, there are five subsections, including the predicted model parameters, the forward data, the observed data, the two covariance matrices, and updated equations, i.e., equations (\ref{mn+1}) and (\ref{cn+1}). These white circles represent random variables, i.e. four model parameters and four observation parameters. It is strait forward to multiple parameters. The predicted model parameters will be predicted by the prediction step, the synthetic data is the theoretical data that will be generated by the forward computation. Note that $d_A$ can be generated by $m_\alpha$, or by $m_\alpha$, $m_\beta$ simultaneously, which relies on whether $d_A$ depends on $m_\beta$. Our model mainly shares correlations through the covariance matrix, which will be generated by the unscented transformation. If $C^{m_\alpha d_B} = 0$, it means that the parameters $m_alpha$ and $d_B$ are not correlated and if $C^{m_\alpha d_B}$, $C^{m_\beta d_A}$, $C^{d_{AB}}$ are all 0, the joint inversion will degenerate to an individual inversion of the two physical processes. 

Although the Kalman inversion does not need to optimize (minimize) an objective function, we use the squared error cost function to purely measure the convergence of the predicted model vector $\ta$
\begin{eqnarray}
\label{eqmm}
    \Phi(\ta) = \frac{1}{2}\left\|\Sigma_{\eta}^{-\frac{1}{2}}(\y-\G(\mu))\right\|^{2},
    \end{eqnarray}
where observed noise $\eta =[\alpha\eta_{A},\beta\eta_{B},\gamma\eta_{C},...]^T$. Given a custom weight is imposed on the data, thus the objective function is defined by the linear combination 
of weighted misfits, thus taking the form:
\begin{eqnarray}
\label{eqrs}
\Phi(\ta)=\alpha\Phi_{A}+\beta\Phi_{B}+\gamma\phi_{C}+....
\end{eqnarray}

\section{Results}
\begin{figure}[htp]
    \noindent\includegraphics[width=\textwidth]{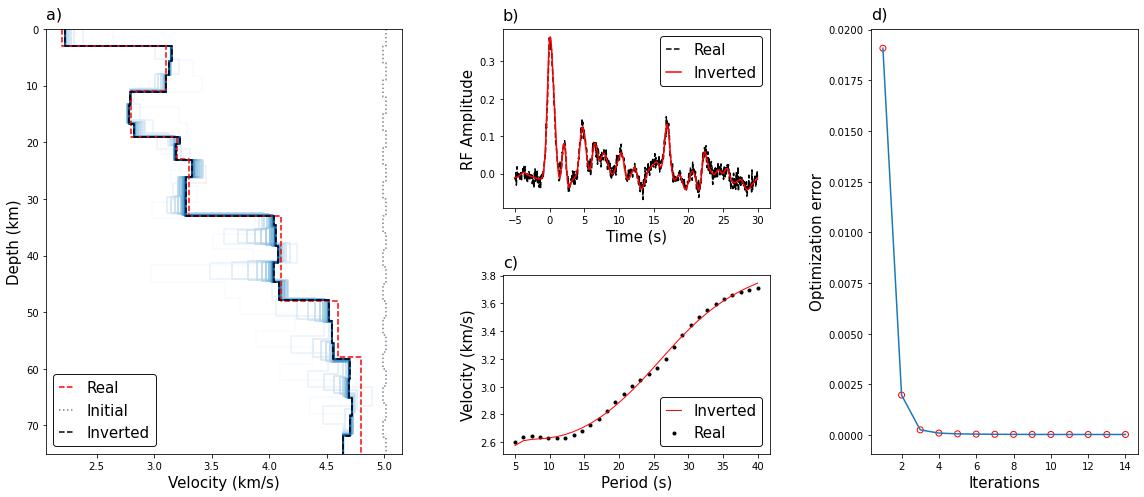}
    \caption{MUKI on the simulated RF and SWD (Phase velocity) data set of the layered model. (a) The inverted model obtained by MUKI runs 6.01 s. The synthesized true velocity model is plotted as a red line, the color from light blue to dark blue shows the evolution of the models through each iteration. (b) Simulated RF data with the Gaussian random noise (black dashed line) and the inversion result (red line). (c) Simulated SWD data with Gaussian random noise and the inversion result. (d) The objective function value at each iteration for the joint inversion using the MUKI.}
    \label{a2}
\end{figure}
Joint inversion of receiver functions (RFs) and surface wave dispersion (SWD) is well-established in geophysics and has been successfully applied with gradient-based methods and Bayesian sampling methods to recover the structure of the earth’s crust and upper mantle. Thus, we use the joint inversion of RFs (only use radial receiver function in this paper) and SWD as a benchmark for our newly developed MUKI to verify its effectiveness. Although it is a general inversion framework for joint inversion, we set the number of observations as 2, i.e., $d_A$ and $d_B$ are RF and SWD data respectively. The model is set up as a horizontally layered model beneath the receivers with the lowest layer a half space. The model consists of 4 parameters, i.e., $m=[v_p,v_s,\rho,h]$, where $v_p,v_s,\rho$ and $h$ are respectively the P wave velocity, S wave velocity, density and thickness at each given layer. In the joint inversion, $v_p$ and $\rho$ can be determined by $v_s$ with empirical relations~\cite{Brocher2005}, which can also be found in Supplementary Text S3. Due to the weak sensitivity of P wave velocity and density to data, we just invert the S wave velocity and thickness of the model.

We set thickness and S wave velocity value in each layer are both variables so that their posterior distributions can be estimated through equations~\ref{mn+1} and \ref{cn+1}. In the joint inversion of RFs and SWD, the mean of the model parameter is initialized as $m=[\mu_vs,\mu_thickness ]^T$.

We demonstrate the effectiveness and efficiency of the MUKI using several tests to invert synthetic and real data of RF(s) and SWD. Throughout all applications, we focus on MUKI. Some comparisons with the UKI, MCMC~\cite{Bodin2012}, and gradient-based methods for the joint inversion of RFs and SWD (CPS,\citeA{Herrmann2013}) are also considered.

\subsection{Synthetic Example}
\label{Synthetic Example}
We use two numerical velocity models to evaluate our algorithm, including \text { i) } Layered model, which consists of 8 horizontal layers with a low S wave velocity layer in the crust and a sharp velocity gradient at the Moho. The model is modified from the 6 layers model referring to~\citeA{Bodin2012a},
\text { ii) } GAr1 model, which can be viewed as a smoother version of model i~\cite{Sambridge1999}. We use the matrix-propagation method to synthesize the seismograms and then obtain the RFs and SWD data~\cite{Haskell1953,Herrmann2013}. Specially, we use the water-level deconvolution method to calculate the RF with a water-level factor of 0.001. Following the work of~\citeA{Bodin2012}, we assume the RF or SWD noise is drawn from the multivariate Gaussian distribution and can be parameterized with two parameters. For SWD, the noise is generated by a diagonal covariance matrix (i.e., n $\sigma_{true}^{SWD}=0.012$ and $r_{true}^{SWD} = 0$); while RF, the noise is generated with an exponential correlation law
with values $\sigma_{true}^{RF} = 0.005$ and $r_{true}^{RF} = 0.92$. 

To initialize MTUKI, we set that the initial model consists of 25 layers, and the means of the thickness of the first seventh layers are all 2km, so the $\mu_{thk}$ can be set as $\mu_{thk}=[2, 2, 2, 2, 2, 2, 2, 3,..., 3]$.  Considering the covariance of the parameter can be iterated updated in the inversion process as equation (\ref{cn+1}), we choose $\mean_0=[\mean_{vs},\mean_{thk}], C_0 = 0.001\mathbb{I}$ and initial MUKI at $m_0 \sim \mathcal{N}[\mu_0,C_0]$, where $\mu_0$ is the prior mean and $C_0$ is an uninformative prior covariance matrix. Note that $\mean_0=[\mean_{vs},\mean_{thk}]$, and the setting of $\mean_{vs}$ will be discussed later.

Figure~\ref{a1} and Figure~\ref{a2} show the inversion processes in the 8-layer model. In Figure~\ref{a1}, it can be seen that the algorithm can adaptively adjust the model parameters~(i.e.velocity and thickness of each layer) (Figure~\ref{a1}a) and fits the observations well~(Figure \ref{a1}b,c). The convergence of the parameter vector is shown in Figure ~\ref{a2}d, Figure~\ref{a2} shows the profiles of the mean and standard deviation (std) of Vs, it can be observed that this method can converge to an "optimal" solution with only $O (10)$ iterations (while CPS need 25 iterations). Compared with the MCMC method (Figure~\ref{a2}b) and gradient method (Figure~\ref{a2}c), our method can obtain an accurate model at least above 60 km with a shorter CPU time (6.01 s/6866.91 s/20 s) on the same computing environment (Intel Core i7-10875H). One detail that must be considered is that our method considers both the thickness and S wave velocity are variables, so the depth of our inversion result is usually different from the initial model setting. Due to the CPS method cannot access the error information directly, we follow the work of \citeA{liGeodynamicProcessesContinental2020} to regenerate theoretical RF and SWD by adding Gaussian random noise as described above, then we obtain the 110 data sets (one RF and one SWD) and then performed the joint inversion respectively to obtain 110 velocity models. After the statistical analysis, we show the results in \ref{a2}c. Since CPS performs the joint inversion using a damped least squares algorithm~\cite{Julia2000} that contains a priori smoothness constraint on the velocity of adjacent layers, thus the results may be smoother than the reference model. 

\begin{figure}[htp]
    \noindent\includegraphics[width=\textwidth]{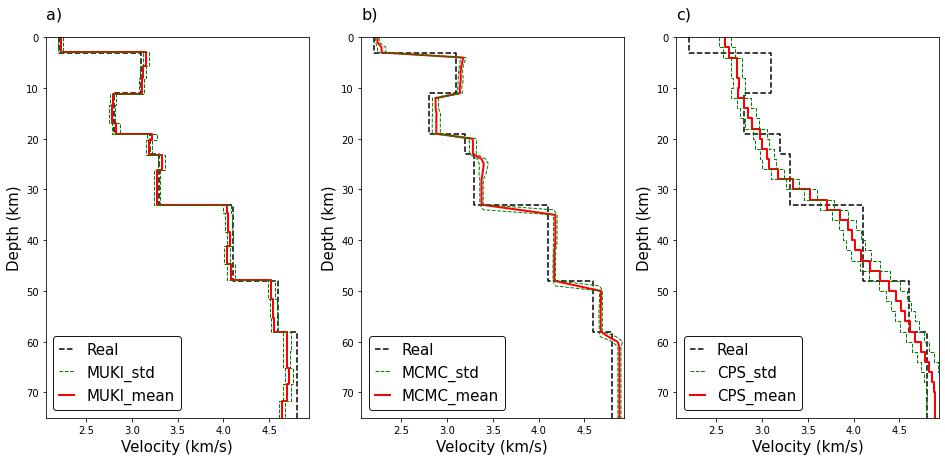}
    \caption{Comparison of MUKI with the MCMC (Bay hunter) method and gradient-based method (CPS) on an 8-layer model. (a) The posterior probability distribution for the S wave velocity at each layer using MUKI, which runs 6.01s. (b) The posterior probability distribution for the S wave velocity at each layer using MCMC (Bay hunter) which obtain 196591 models from 7 chains and runs 6866.91 s. (c) The joint inversion results were obtained by the CPS.}
    \label{a3}
\end{figure}

The comparisons between the joint inversion result with the results of individual RF and SWD inversions are shown in Figure S1, which shows that the joint inversion can obtain the best parameter estimation against the individual inversions of SWD and RF.

Then we design several tests to investigate the influence of the initial Vs on our result (Figure S2). In these tests, the initial $\mu_{thk}$ are identical but $\mu_{vs}$ are different. This numerical experiment demonstrates that discontinuities with sharp speed changes (i.e., Moho interface) can be effectively recovered with different initial $\mu_{vs}$.

\begin{figure}[htp]
    \noindent\includegraphics[width=\textwidth]{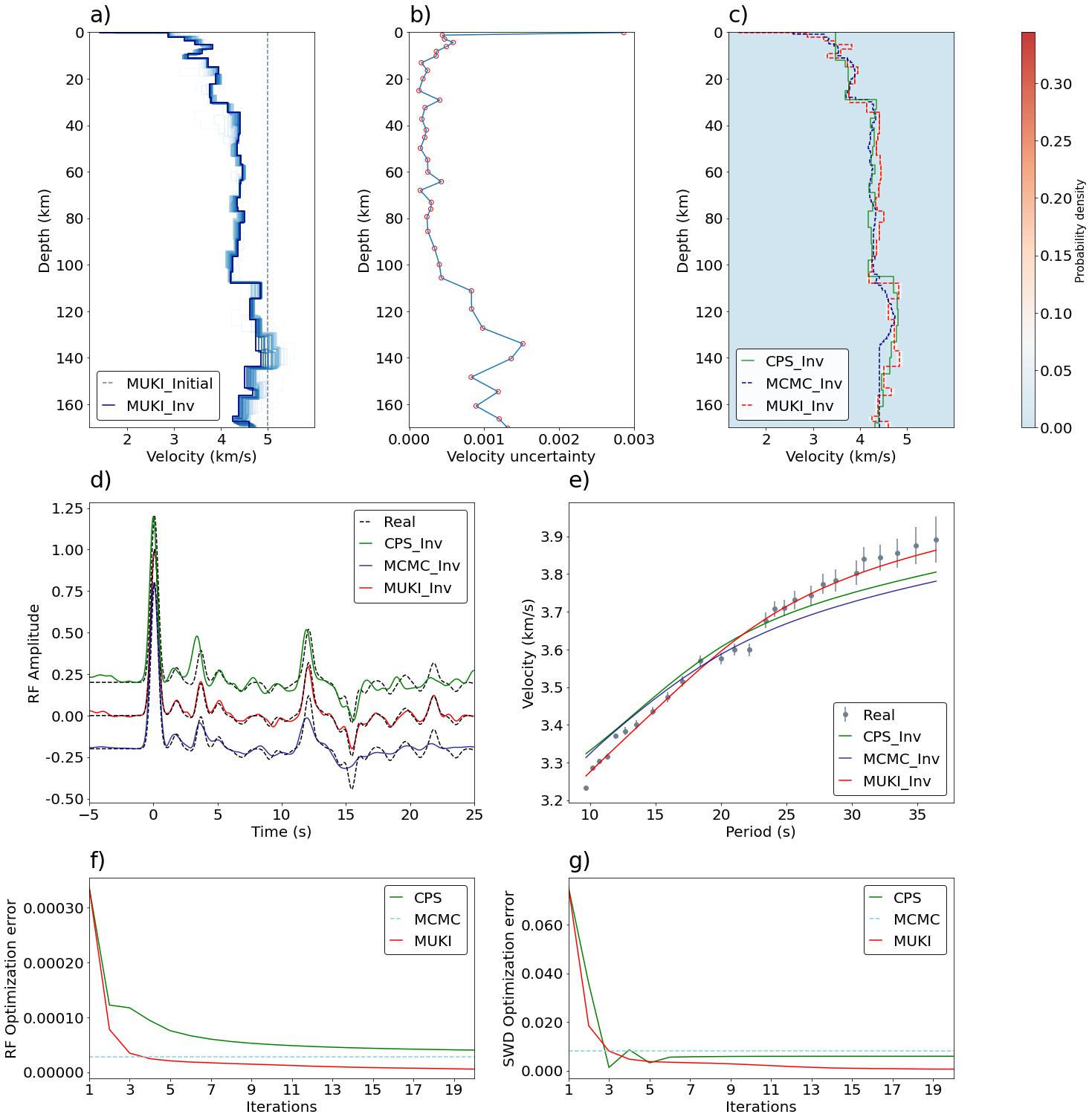}
    \caption{Joint inversion of field data for the station KIGAM. (a) Evolution of the inverted models obtained by MUKI. (b) The uncertainty for the S wave velocity of each layer, which can be obtained from the diagonal of the covariance matrix. (c) The inversion results were obtained by the three methods and the posterior probability distribution was obtained by MUKI.  (d) The model fitnesses for the receiver function with three methods. To better show three inversion results from one RF in one figure, each one is shifted by an offset of -0.25, 0, 0.25, respectively. (e) The model fitnesses for the dispersion data with three methods. (f)-(g) are separately the optimization error of RF and SWD using three methods (MCMC is not an iterative algorithm, we show the final model (mean model) error with a green dashed line).}
    \label{a4}
\end{figure}

\subsection{Field Case}
Our algorithm is further tested using a real data set from the KIGAM station collected by Seoul National University, which is located at $(35.00^\circ N,126.249^\circ E)$. The dispersion measurements and receiver functions at station KIGAM have been obtained and performed well~\cite{Herrmann2013}. We performed experiments to compare the MUKI method with the other two methods. The RF was calculated using the time domain iterative deconvolution procedure~\cite{Ligorria1999}, which is more stable in the presence of noisy data than the frequency domain method. The RF has a Gaussian factor alpha=2.5 and the period of SWD (Phase velocity) varies from 10 s to 37 s. The initial mean of thickness and S wave velocity are shown in Figure~\ref{a4}a, which is the same as the setting in CPS.

The joint inversion results for the station KIGAM are shown in Figure~\ref{a4}. Figure~\ref{a4}a shows that it converges after twenty iterations, which are drawn from light blue line to dark blue line, and can invert the distribution of thickness and S wave velocity in each iteration. Since we only use the surface wave dispersion of less than 40 s, the uncertainty of the inversion results will increase rapidly with depth when the depth exceeds 80 km. To illustrate this, we have plotted in Figure~\ref{a4}b the uncertainty for the S wave velocity of each layer. As the depth increases, the maximum probability density of S wave velocity at each depth will decrease, which means an increase in uncertainty. As RF has strong non-uniqueness for the shallow S wave velocity, leading to large uncertainties. The solutions obtained by the MUKI (red dashed line), CPS (green line), and MCMC (blue line) are shown in Figure \ref{a4}c, respectively. The probability density of the posterior distribution of Vs is plotted as the base map. In Figures 4d and 4e, it can be found that the structure recovered by MUKI can fit the observed RF and SWD best, which can also be confirmed by the optimization error in Figures \ref{a4}f, and \ref{a4}g. These results verify that the model recovered by our method in Figure \ref{a4}b is more reliable, and the discontinuous interface around 32-34 km   (Moho layer \cite{Chang2007}) can be recovered well.   

\section{Conclusions}
In this study, we develop a new derivative-free joint inversion framework and evaluate it by joint inversion of RF and SWD. In both synthetic and real data set tests, the newly developed inversion framework demonstrates its powerful ability in terms of at least three aspects: a) Due to it is derivative-free inversion framework, it honors good flexibility in inverting multi-physical geophysical data; b) It can effectively obtain uncertainty of the solution by Gaussian approximation; c) It has a fast convergence rate compared to traditional methods.

\section{Open Research}
RF data and the SWD data from the station KIGAM can be accessed through the CPS tutorial \url{(http://www.eas.slu.edu/eqc/eqc_cps/TUTORIAL/STRUCT/index.html)}.

\acknowledgments

This research was jointly supported by the National Key R$\&$D Program of China (grant 2016YFC0600402), the Strategic Priority Research Program (B) of the Chinese Academy of Sciences (grant XDB18000000), and the National Natural Science Foundation of China (grants 41374063 and 41874065).

We appreciate Daniel Z. Huang for his outstanding work about UKI. Thanks to Ran You for his great helps with UKF. We are grateful to R.B. Herrmann and Jennifer Dreiling for providing the CPS software and MCMC code used in this study. Thanks to Sicheng Zuo for revising the manuscript. Thanks to Wentao Li and Yifan Lu for the valuable discussion about the manuscript. Thanks to colleagues in our lab for helpful discussions on the manuscript.

\bibliography{ref}
\end{document}